\documentclass[reprint,amsmath, superscriptaddress, amssymb, aps,longbibliography,prl,showkeys]{revtex4-1}

\usepackage{graphicx}
\usepackage{dcolumn}
\usepackage{bm}
\usepackage{nccmath}

\usepackage[T1]{fontenc}
\usepackage[utf8]{inputenc}
\usepackage{physics}

\usepackage{xcolor}

\newcommand\hl[1]{%
  \bgroup
  \hskip0pt\color{blue}%
  #1%
  \egroup
}

\newcommand\hlold[1]{%
  \bgroup
  \hskip0pt\color{red}%
  #1%
  \egroup
}

\DeclareMathAlphabet{\pazocal}{OMS}{zplm}{m}{n}

\begin{document}

\title{Excitonic thermalization bottleneck in twisted TMD heterostructures}

\author{Giuseppe Meneghini}
\email{giuseppe.meneghini@physik.uni-marburg.de}
\affiliation{%
 Department of Physics, Philipps University of Marburg, 35037 Marburg, Germany}%

\author{Samuel Brem}
\affiliation{%
 Department of Physics, Philipps University of Marburg, 35037 Marburg, Germany}%

\author{Ermin Malic}
\affiliation{%
 Department of Physics, Philipps University of Marburg, 35037 Marburg, Germany}%

\date{\today}

\begin{abstract}
\textbf{Abstract:} 
Twisted van der Waals heterostructures show an intriguing interface exciton physics including hybridization effects and emergence of moir\'e potentials. Recent experiments have revealed that moir\'e-trapped excitons exhibit a remarkable dynamics, where excited states show lifetimes that are several orders of magnitude longer than those in monolayers.  The origin of this behaviour is still under debate. Based on a microscopic many-particle approach, we investigate the phonon-driven relaxation cascade of non-equilibrium moir\'e excitons in the exemplary MoSe$_2$-WSe$_2$ heterostructure. We track the exciton relaxation pathway across different moir\'e mini-bands and identify the phonon-scattering channels assisting the spatial redistribution of excitons into low-energy pockets of the moir\'e potential. We unravel a phonon bottleneck in the flat band structure at low twist angles preventing excitons to fully thermalize into the lowest state explaining the measured enhanced emission intensity of excited moir\'e excitons. Overall, our work provides important insights into exciton relaxation dynamics in flatband exciton materials.
\end{abstract}


\maketitle

Van der Waals heterostructures consisting of monolayers of transition metal dichalcogenides (TMDs) have been intensively studied in the past years \cite{liu2016van,li2018engineering,wilson2021excitons,hagel2021exciton}. 
The type-II heterostructure facilitates 
the emergence of interlayer excitons, where the Coulomb-bound electrons and holes are spatially separated in opposite layers \cite{rivera2015observation,enalim2019restoring,merkl2019ultrafast,hagel2021exciton}. These states are characterized by a long lifetime and exhibit a permanent out-of-plane dipole moment,  making them promising for technological applications \cite{ross2017interlayer,jin2019identification,tan2021layer}. In addition, the presence of a strong tunneling between the layers allows the existence of layer-hybridized states, in which the intra- and interlayer nature can be controlled by electrical fields \cite{erkensten2023electrically,tagarelli2023electrical}. Hybrid excitons have been shown to play a key role for the charge transfer process in these materials \cite{jin2018ultrafast,meneghini2022ultrafast,schmitt2022formation} as well as for the transport behaviour \cite{tagarelli2023electrical}.
The possibility of introducing a twist angle between  two vertically stacked TMD layers has opened the door to fascinating  physical phenomena that are governed by periodic moir\'e potentials. \cite{mak2022semiconductor,andrei2021marvels}. In the range of small twist angles ($\leq$ 2$^\circ$), long-lived moir\'e-trapped exciton states have been demonstrated resulting in a multi-peaked structure in photoluminescence (PL) spectra \cite{tran2019evidence,brem2020tunable,huang2022excitons}.  The microscopic origin of the moir\'e-peak series can be traced back to the radiative recombination of excitons located  in different moir\'e sub-bands. The latter are flat at small twist angles reflecting their localization in real space. 

\begin{figure}[t!]
  \centering
  \includegraphics[width=\columnwidth]{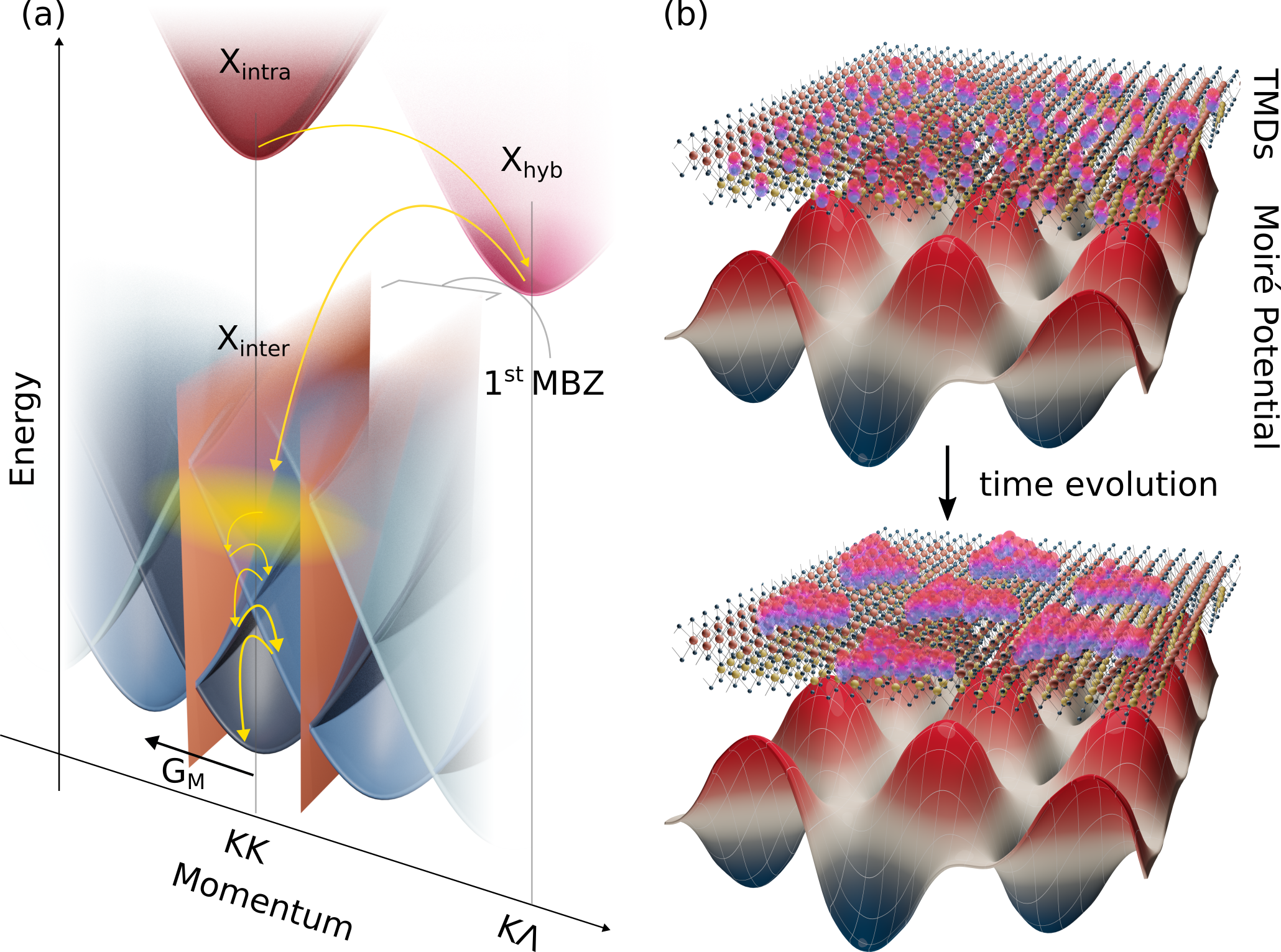}
  \caption{Sketch of exciton dynamics in a twisted TMD heterostructure. (a) After optical excitation of  intralayer excitons (X$_{intra}$) in one of the layers, exciton population relaxes to the  energetically lowest states (interlayer excitons X$_{inter}$ in the case of MoSe$_2$-WSe$_2$ investigated here) via momentum-dark hybrid excitons X$_{hyb}$. We focus here on microscopic modelling the relaxation cascade of hot interlayer excitons. We depict the exemplary case of parabolic bands, in which a new periodicity (G$_M$ reciprocal lattice moir\'e vector) arises. In the case of small twist angles, there is, in addition to the new periodicity, a change in the band curvature, resulting in flat bands. This modifies drastically the allowed scattering channels. (b) Exciton relaxation in momentum space is reflected by the change of exciton localization in real space: the thermalization process brings the exciton population (purple dots) from a delocalized phase to the most favourable configuration of trapped states.}
  \label{fig:schematic}
\end{figure}

A thorough microscopic understanding of moir\'e exciton physics is highly interesting for fundamental science and also of key importance for  the technological application potential of van der Waals heterostructures \cite{choi2017recent,yu2017moire,tran2020moire}. The strong localization and the non-trivial band topology give rise to remarkable quantum many-body effects \cite{shimazaki2020strongly,wang2020correlated}, varying from spin-liquid states \cite{wu2018hubbard,kiese2022tmds}, Mott insulating states \cite{tang2020simulation,regan2020mott,li2021continuous}, and even superconductivity \cite{shi2015superconductivity,qiu2021recent}. 
Several models have studied these exotic states, applying the common approach of mapping the system to Hubbard-like models \cite{wu2018hubbard,pan2020quantum,tang2020simulation,gotting2022moire,xu2022tunable}.
The starting point and common assumption for these models is that all carriers occupy the lowest mini-band/localized orbital.  However, in typical experiments, the excitation with a laser creates a population of hot excitons that first have to dissipate their thermal energy to reach the ground state. A microscopic modeling of the relaxation cascade of hot excitons along the moir\'e sub-band structures that consists of many flat bands is a  challenging task. Experimental observations hint at the presence of non-thermal exciton distributions resulting in long-lived excited states. Several experimental studies have demonstrated that their lifetime is on the order of nanoseconds in marginally twisted van der Waals heterostructures \cite{choi2021twist,tran2019evidence,li2021interlayer}. PL spectra show that the optical response of excited states can be much brighter compared to the ground state, indicating a strong non-equilibrium exciton distribution \cite{tran2019evidence}.
Although these experiments clearly hint at an unconventional relaxation dynamics in presence of a moir\'e potential, there is still little known about the underlying microscopic processes. 

In this work, we study the phonon-driven relaxation cascade of hot interlayer excitons in  the exemplary twisted  MoSe$_2$-WSe$_2$ heterostructure. We focus on the low twist angle regime characterized by moir\'e-trapped excitons and a flat moir\'e sub-band structure \cite{tran2019evidence,jin2019observation,brem2020tunable,brem2023bosonic}.  Our study is based on a microscopic many-particle approach and allows us to track the phonon-driven relaxation pathway of excitons from an initial hot exciton distribution towards the ground state.
We conduct a temperature-, momentum- and time-dependent study of the exciton relaxation dynamics identifying the presence of a pronounced relaxation bottleneck for small twist angles ($\simeq$ 1$^\circ$) and low temperatures ($<$ 100 K), in particular preventing excitons to efficiently scatter from the first and second excited states to the ground state. We calculate the time-dependent PL and show that the bottleneck effect manifests in a significant occupation of  excited states and results in their unexpectedly high emission - in excellent qualitative agreement with observations in experimental PL spectra \cite{tran2019evidence}.

\begin{figure*}[t!]
  \centering
  \includegraphics[width=\textwidth]{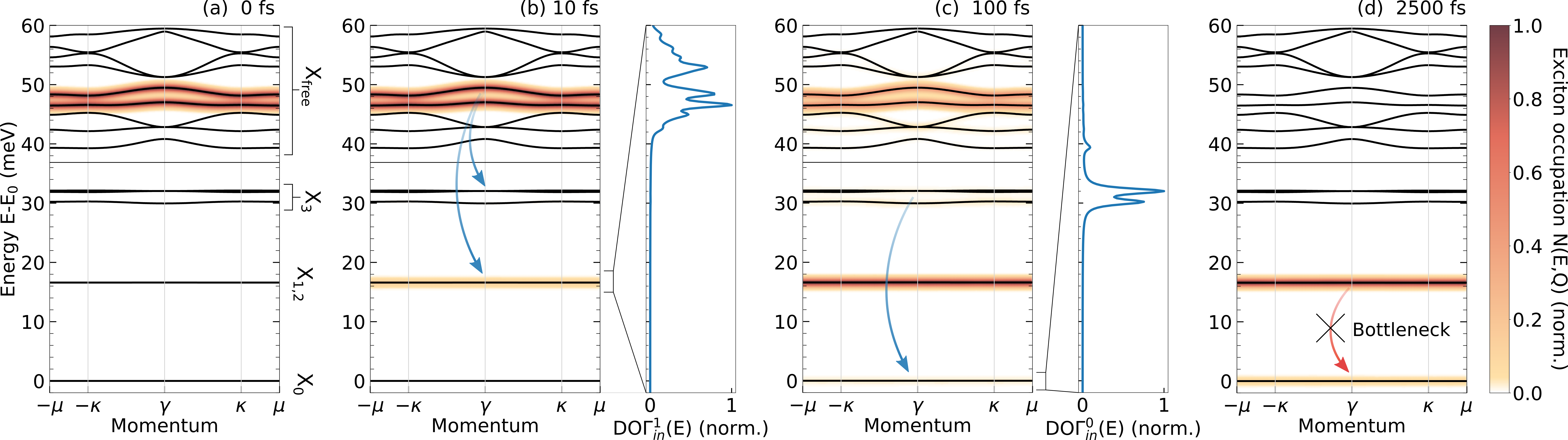}
  \caption{Interlayer exciton energy landscape of the MoSe$_2$-WSe$_2$ heterostructure (with a 1$^\circ$ twist angle) consisting of  bound states (X$_{0,1,2}$), intermediate states (X$_3$), and free states (X$_{free}$). All  energies are plotted with respect to the ground state  (E$_0$). We show superimposed on the bands the energy- and momentum-resolved exciton occupation (red-orange shaded) at subsequent steps of the dynamics at 40 K: (a) We start at 0 fs with an initial hot distribution of excitons (created through scattering from optically excited intralayer exciton states, cf. Fig. \ref{fig:schematic}a). (b) At an early stage of the dynamics (10 fs), the emission of optical phonon drives the population predominantly to the first degenerate excited states X$_{1,2}$. (c) X$_0$ is occurring at a much slower speed, driven by the filling of intermediate states X$_3$ due to the scattering with acoustic phonons (100 fs). This different energy dependence in the relaxation becomes clear by analyzing the density of in-scattering states 
  $\mathrm{DO\Gamma}^\eta_{\text{in}}(E)$
   (see the text for the definition) for X$_0$ and X$_{1,2}$, illustrating which energy window contributes the most to the increase of the exciton population of these lower states. (d) In the final stage of the dynamics on the timescale of a few ps, we observe a bottleneck effect, i.e.  the scattering to the ground state is almost completely suppressed. This results in a strong out-of-equilibrium exciton distribution, where excited states show a higher occupation than the ground state.}
  \label{fig:dynamics}
\end{figure*}

\textbf{Theoretical framework:} To microscopically understand the exciton dynamics in twisted van der Waals heterostructures, we first need to derive the key equations describing the motion of excitons in presence of a periodic moir\'e potential. We start from the general Hamilton operator for TMD bilayers including the energy of intra- and interlayer excitons and their interaction with phonons using a valley-local approach \cite{ovesen2019interlayer,brem2020hybridized} yielding $H = \sum_{\mu}\mathcal{E}^\mu_{ \bf Q} X^{\mu \dagger}_{ \bf Q} X^\mu_{ \bf Q} + \sum_{\mu \nu j {\bf q Q}} D^{\mu \nu}_{j {\bf q Q}} X^{\nu \dagger}_{{\bf Q+q}} X^\mu _{{\bf Q}} b_{j {\bf q}} + h.c.$ with  $\mathcal{E}^\mu_{ \bf Q}$ as the free excitonic energy, $D^{\mu \nu}_{j {\bf q Q}}$ as the exciton-phonon matrix element, and $b_{j {\bf q}}$ as the phonon annihilation operator labeled by the phonon mode $j$ and the momentum transfer ${\bf q}$. The Hamiltonian includes 
$X^{\mu (\dagger)}_{ \bf Q}$  as the exciton creation/annihilation operator with ${\bf Q}$ as the center-of-mass momentum of the electron-hole pair  and  $\mu$ as the compound index containing the  excitonic quantum numbers $\mu = n,\xi_e,\xi_h,l_e,l_h$, where $n$ denotes the series of Rydberg-like states, $\xi_{e/h}$  the electron and hole valley index, and  $l_{e/h}$ the layer index. 
Since our goal is to understand the exciton  relaxation cascade in presence of a  moir\'e potential, we focus on the lowest interlayer exciton series of states that are known to be located at the K point in the MoSe$_2$-WSe$_2$ heterostructure investigated here. Note that one can neglect the hybridization of intra- and interlayer exciton states as here the wavefunction overlap is known to be small, while the energetic detuning is large \cite{cappelluti2013tight, gillen2018interlayer,merkl2020twist, brem2020hybridized}.

The presence of a twist angle can be included  in terms of an effective potential, arising from the spatially dependent atomic local alignment \cite{brem2020tunable}, denoted as moir\'e potential $V_M = \sum_{\bf Q g} \mathcal{M}_{\bf g} X^{\dagger}_{ \bf Q +g } X_{ \bf Q }$. Here, we have introduced ${\bf g} = s_1 {\bf G^M_1} + s_2 {\bf G^M_2}$ with ${\bf G^M_{1/2}}$ as reciprocal moir\'e lattice vectors and $s_{1/2}$ as integers and  $\mathcal{M}_{\bf g}$ refers to the effective exciton potential generated by the local displacement of the two twisted layers (more  details can be found in the SI). By changing into a moir\'e exciton basis and introducing  creation and annihilation operators for moir\'e excitons $Y^{\eta }_{\bf Q} = \sum_{\bf g} \omega^{\eta}_{\bf g}({\bf Q}) X_{\bf Q + g}$ with the mixing coefficient $\omega^{\eta}_{\bf g}({\bf Q})$ corresponding to the Bloch wave function, we transform the Hamilton operator described above into \cite{brem2020tunable}
\begin{equation*}
    \label{eq:hamiltonian}
    \tilde{H} = \sum_{\eta}E^\eta_{ \bf Q} Y^{\eta \dagger}_{ \bf Q} Y^\eta_{ \bf Q} + \hspace{-8pt}\sum_{\eta \xi j, \bf Q Q' g}\hspace{-8pt}  \Tilde{\mathcal{D}}^{\eta \xi j}_{{\bf Q Q' g }} Y^{\xi \dagger}_{{\bf Q'}} Y^\eta _{{\bf Q}} b^j_{{\bf Q'-Q+  g}} + h.c.
\end{equation*}
with $\Tilde{\mathcal{D}}^{\eta \xi j}_{{\bf Q Q' g }}$ as the exciton-phonon coupling tensor in the new basis containing the overlap of initial and final moir\'e states.
Now, we use this moir\'e exciton Hamiltonian to solve the Heisenberg equation of motion for the exciton occupation $N^\eta_{\bf Q} = \big<Y^{\eta \dagger}_{ \bf Q} Y^\eta_{ \bf Q}\big>$. Within the second-order Born-Markov approximation, we obtain a Boltzmann scattering equation describing the phonon-mediated relaxation dynamics of the exciton occupation \cite{ thranhardt2000quantum,haug2009quantum,selig2018dark,brem2018exciton}  
\begin{equation}
    \label{eq:time_evolution}
    \Dot{N}^\eta_{\bf Q}(t) = \sum_{\xi {\bf Q'}} \left[\mathcal{W}^{\xi \eta}_{ \bf Q' Q} N^\xi_{ \bf Q'}(t) + \mathcal{W}^{ \eta \xi}_{ \bf Q Q'} N^\eta_{ \bf Q}(t)\right]
\end{equation}  
with the scattering tensor $\mathcal{W}^{\eta \xi}_{ \bf  Q Q'}$ containing the microscopically calculated transition rates between different exciton states driven by emission and absorption of optical and acoustic phonons. Here, we include a collisional broadening (third-order terms) to take into account a self-consistent temperature-dependent broadening that softens the energy conservation (more details can be found in the supplementary material). Note that we focus on the low-density, regime where exciton-exciton scattering (such as Auger-type processes) can be neglected \cite{erkensten21}.  
This equation allows us to track the relaxation cascade of excitons from an initially created hot distribution towards the ground state resolved in time and momentum. 

In the following, we focus on the analysis of the time evolution of the  momentum-integrated exciton occupation $N^\eta(t) = \sum_{\bf Q} N^\eta_{\bf Q}(t)$ and of the energy- and momentum-resolved occupation $N(E,{\bf Q}, t) = \sum_{\eta} N^\eta_{\bf Q}(t) \delta(E - E^\eta_{\bf Q})$. In addition, we introduce the density of in-scattering states $\mathrm{DO\Gamma}^\eta_{\text{in}} (E,t)= \sum_{\xi {\bf Q'}} \mathcal{W}^{\xi\eta }_{ \bf  Q', Q = 0} (t) \delta\left( E - E^\xi_{\bf Q'}\right)$, quantifying the density of states contributing the most to the influx of excitons to a specific state $\eta$. To perform a quantitative analysis and be able to compare our predictions with experiments, we compute the time- and energy-dependent photoluminescence intensity $I_{PL}(E,t)$ using \cite{rossi2002theory}
\begin{equation}
\label{eq-pl}
    I_{PL}(E,t) \propto \sum_{\eta,\sigma} \gamma^\eta_{\sigma} N^\eta_{\bf Q = 0}(t) \mathcal{L}_{\Gamma^\eta}(E^\eta_{\bf Q = 0} - E)  
\end{equation}
with $\gamma^\eta_{\sigma}$ containing the optical selection rules for moir\'e excitons and $\mathcal{L}_{\Gamma^\eta}$ denoting the Cauchy/Lorentz distribution. The width $\Gamma^\eta = \sum_{\xi,{\bf Q'}}\mathcal{W}^{\eta \xi}_{\bf Q = 0, Q'}$  is determined by evaluating the exciton scattering tensor at  ${\bf {Q}}=0$.  To be able to quantify the efficiency of the relaxation process,  we introduce the relaxation time $\tau_{\eta \rightarrow \xi} = 1/\Gamma^{\eta \rightarrow \xi}_{\bf Q = 0} = 1/ \sum_{{\bf Q'}}\mathcal{W}^{\eta \xi}_{\bf Q = 0, Q'}$ desribing the scattering efficiency  from the state $\eta$ to $\xi$. Further details on the theoretical approach can be found in the supplementary material. \\

\textbf{Relaxation bottleneck effect:}
We now apply the theoretical framework described above and  numerically evaluate Eq. (\ref{eq:time_evolution}) for the specific case of the twisted MoSe$_2$-WSe$_2$ heterostructure, focusing on small twist angles ($\simeq$ 1$^\circ$) and low temperatures ($\simeq$ 40 K). Our aim is to study the relaxation cascade of injected hot interlayer excitons in  presence of a periodic moir\'e potential (Fig. \ref{fig:schematic}a). Typically, intralayer excitons are optically excited in one of the two layers and they then scatter down to the energetically lower interlayer exciton states. In this work, we focus on the final stage of this dynamics, i.e. the relaxation cascade within the series of interlayer moir\'e mini-bands. Atomic reconstruction is typically more impactful at smaller twist angles and has thus been neglected here \cite{weston2020atomic}.

We start our analysis  by investigating how the presence of a moir\'e potential modifies the interlayer exciton landscape in the twisted MoSe$_2$-WSe$_2$ heterostructure. 
The energy landscape is obtained by solving the eigenvalue problem including the angle-dependent moir\'e potential. This gives us access to the new eigenenergies of the system, $E^\eta_{\bf Q}$, shown in Fig. \ref{fig:dynamics}. We can  distinguish nearly flat bound states ($X_{0,1,2,3}$), characterized by wavefunctions localized around minima of the moir\'e potential, and free states ($X_{free}$) which show a more delocalized wavefunction \cite{brem2020tunable}. After having calculated the exciton energy landscape, our first goal is to reveal the microscopic origin of the measured long lifetimes of excited moir\'e state \cite{choi2021twist,tan2022signature}. To address this question, we analyse the phonon-driven transition rates and the relaxation time from the excited states (X$_{1,2}$) to the ground state X$_0$. 

We first investigate the time evolution of the exciton occupation $N(E,{\bf Q})$ that we plot superimposed on the moir\'e exciton band structure in Fig. \ref{fig:dynamics}. This way we can directly track the relaxation pathway of excitons. We start with an initial exciton distribution in the energy range of free states, specifically around 40-50 meV away from the ground interlayer exciton state (Fig. \ref{fig:dynamics}(a)). The subsequent thermalization of moir\'e excitons can be described in terms of two competing processes, driven by emission of optical and acoustic phonons, respectively. Scattering with acoustic phonons, characterized by a small transfer of energy and momentum, populates the adjacent energy bands, i.e. the intermediate states X$_3$. The scattering with optical phonons makes excitons dissipate faster and relax further down to the first excited states X$_{1,2}$, cf. the arrows in Fig. \ref{fig:dynamics}(b).

To better understand the relaxation path of excitons, we show in Figs.  \ref{fig:dynamics}(b)-(c) the density of in-scattering states $\mathrm{DO\Gamma}^\eta_{\text{in}}$ for the ground state $\mu=0$ and for the first excited states $\mu=1,2$. This quantity discloses the energy window of initial states contributing the most to the population of the final state $\mu$. Considering the case of the first excited states, we find that the energy window including the free states that have been initially populated contributes the most (Fig. \ref{fig:dynamics}(b)). On the other hand,  the ground state is mostly populated from the in-scattering from the intermediate states X$_3$ (Fig. \ref{fig:dynamics}(c)). As a result, the occupation of X$_{1,2}$ that is driven by emission of optical phonons, occurs on a much faster timescale of a few tens of fs compared to the population of the  ground state on a timescale of hundreds of fs. For the latter to be filled, excitons have first to relax via acoustic phonons to the intermediate states X$_3$. 
Figure \ref{fig:dynamics}(d) illustrates that, interestingly, even for longer times of a few ps, the occupation of the ground state remains clearly lower than the one of the first excited states - in contrast to what we would expect from a thermalized distribution. This indicates the emergence of a pronounced relaxation bottleneck that keeps the exciton occupation out of thermal equilibrium. 

Note that the initial condition chosen for the study is based on the observation of DO$\Gamma^1_{in}$(E). In order to obtain the strongest out-of-equilibrium distribution of excitons, one has to excite in an energy window in which the population is scattering faster to X$_{1,2}$ than to X$_3$.
Exciting in a different energy window changes the quantitative distribution of the excitonic occupation of X$_0$ and X$_{1,2}$ but our key result, i.e. the emergence of a relaxation bottleneck giving rise to a non-thermal exciton distributions at low temperatures, remains unaffected. A more detailed discussion can be found in the SI, where we explicitly show the impact of different initial conditions on the relaxation dynamics of moir\'e excitons.
\begin{figure}[t!]
  \centering
  \includegraphics[width=\columnwidth]{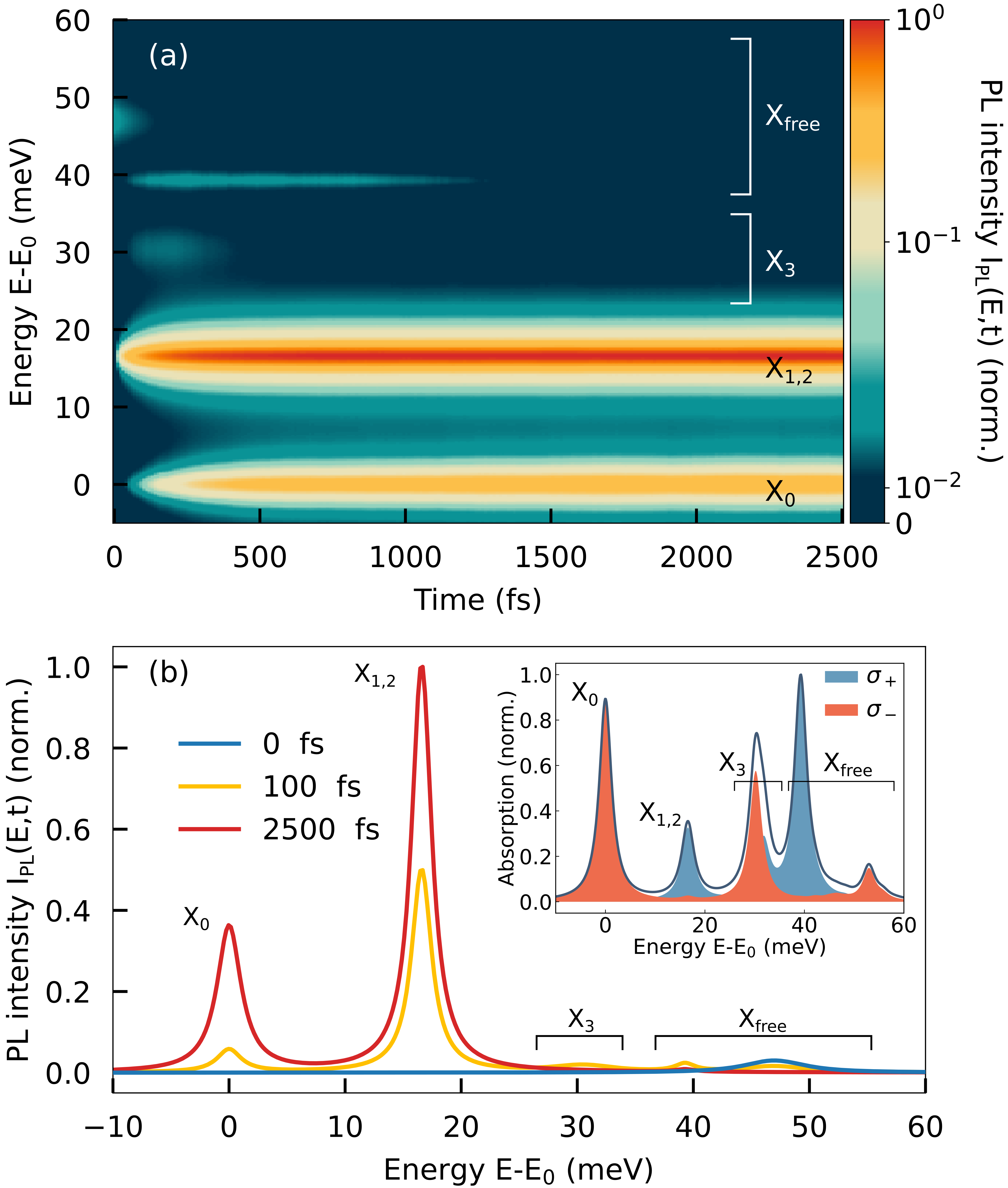}
  \caption{(a) Photoluminescence spectrum as a function of energy and time and (b) at different fixed time cuts. We include the absorption spectrum as inset in (b) to highlight the optical selection rules of different states. Here, the solid grey line shows the total absorption, while red and blue lines denote the contribution of the $\sigma_-$ and $\sigma_+$ circularly polarized light, respectively.}
  \label{fig:PL_spectra}
\end{figure}

To be able to compare our predictions with experiments, we determine the consequences of the predicted non-equilibrium exciton occupation on time-dependent PL spectra  by evaluating Eq. \eqref{eq-pl}. 
We find a clearly higher emission from the first excited states X$_{1,2}$ than from the ground state X$_{0}$ (Fig. \ref{fig:PL_spectra}) reflecting directly the higher occupation of X$_{1,2}$ shown in Fig. \ref{fig:dynamics}(d). This finding is in excellent qualitative agreement with experimental measurements \cite{tran2019evidence}.
According to  Eq. \eqref{eq-pl}, the  PL intensity depends on the exciton occupation in a certain state weighted by its optical matrix element. The latter describes the oscillator strength of the states and can be directly accessed in a linear absorption spectrum (cf. the inset \ref{fig:PL_spectra}(b)).  We find that  the absorption peak of X$_{1,2}$ has half intensity with respect to the ground state X$_0$, and that  only one of the two degenerate states X$_{1,2}$  is optically active \cite{brem2020tunable}. For these reasons, a higher PL peak of  X$_{1,2}$ means that its occupation has to be significantly larger than the one of X$_{0}$, which is only the case for a  highly non-equilibrium exciton distribution emerging as a consequence of a pronounced relaxation bottleneck. \\

\begin{figure}[t!]
  \centering
  \includegraphics[width=\columnwidth]{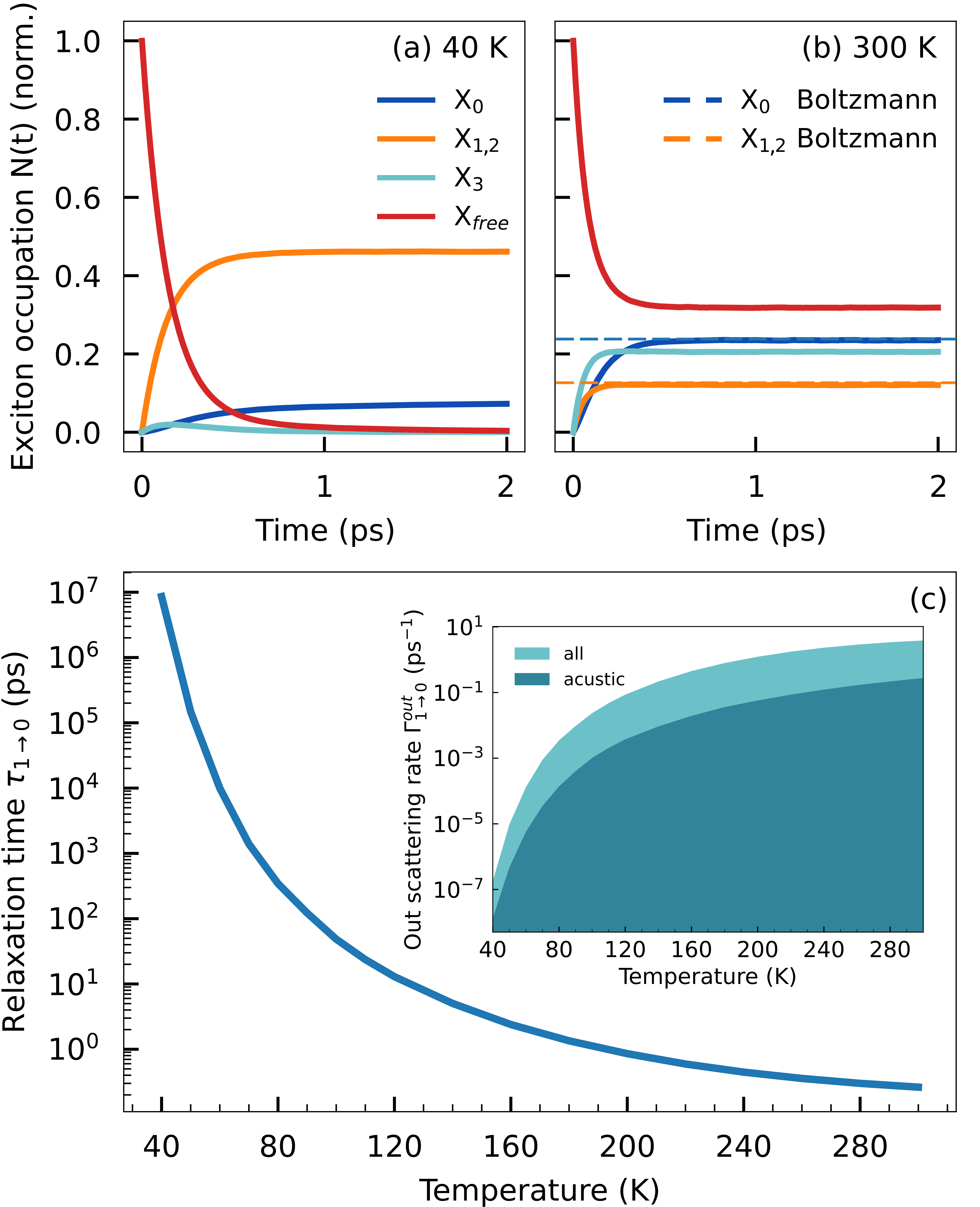}
  \caption{Momentum-integrated exciton dynamics at (a) 40 K and (b) 300 K. At low temperatures, we observe a much larger stationary occupation of the first excited states X$_{1,2}$, highlighting the importance of the relaxation bottleneck leading to a strong deviation from a thermal distribution. In contrast, at higher temperatures,  the exciton occupation clearly relaxes into a Boltzmann distribution (dashed lines). (c) The relaxation time $\tau_{1\rightarrow 0}$ from the first excited state to the ground state as function of temperature, showing that the bottleneck effect becomes significantly strong at temperatures smaller than approximately 100 K leading to recombination times in ns or even $\mu$s range. In the inset, we plot the out-scattering rate $\Gamma^{\text{out}}_{1\rightarrow 0}$, identifying the different contributions of acoustic and optical phonons. }
  \label{fig:tau_scattering}
\end{figure}

\textbf{Temperature dependence of the bottleneck:}
To further characterize the relaxation bottleneck effect, we perform a temperature-dependent study of the exciton relaxation dynamics. A suitable quantity to track the emergence of the bottleneck is the relaxation time from the first excited states to the ground state, $\tau_{1 \rightarrow 0}$. We directly compare  the momentum-integrated time-dependent exciton occupation at 40 K and 300 K, cf. Fig.  \ref{fig:tau_scattering}.  We find  a qualitatively different exciton dynamics: in the low temperature case, we observe a  strong out-of-equilibrium distribution of excitons, with excited states X$_{1,2}$ containing a roughly eight times larger population than the ground state X$_0$ (Fig. \ref{fig:tau_scattering}(a)). This behavior is not seen at room temperature, where the initially hot excitons dissipate all their excess energy, reaching a Boltzmann distribution on a sub-ps timescale (cf. the solid and dashed lines in Fig. \ref{fig:tau_scattering}(b)). This hints at a  temperature-dependent activation of the bottleneck effect and is further confirmed by analysing the temperature-dependent relaxation time $\tau_{1 \rightarrow 0}$, shown in Fig.  \ref{fig:tau_scattering}(b). 
Here, we find a huge variation of several orders of magnitude in $\tau_{1 \rightarrow 0}$: For temperatures lower than 100 K, the  transition from X$_{1,2}$ to X$_0$ becomes drastically slowed down resulting in an extremely large relaxation time $\tau_{1 \rightarrow 0}$ being on a timescale comparable or even longer than the recombination time of interlayer excitons of typically $\simeq 10^2-10^3$ ps \cite{jiang2021interlayer}. 

This temperature study helps us reveal the microscopic origin of the bottleneck effect. In  the inset of Fig. \ref{fig:tau_scattering}(b) we  show the out-scattering rate from  $\Gamma^{out}_{1 \rightarrow 0}$ (that is the inverse of $\tau_{1 \rightarrow 0}$), where we separate the two contributions stemming from the emission of acoustic and optical phonons. We find clearly that  the most efficient contribution to exciton thermalization arises from optical phonons. They are  at least one order of magnitude larger than the scattering with acoustic phonons. This can be understood in terms of exciton energies and center-of-mass momenta. 
The conservation of energy contained in $\Tilde{\mathcal{D}}^{\eta \xi j}_{{\bf Q Q' g }}$, in combination with the flatness of the bands, imposes a strong boundary condition to the available scattering partners for this transition. Given the excitonic flat dispersion for both the initial (X$_{1,2}$) and the final state (X$_0$), the energy conservation plays the key role.
The energy difference between X$_{1,2}$ and X${_0}$ is $\simeq 16$ meV. Acoustic phonons, given their linear dispersion, would require a huge momentum transfer to be able to dissipate this amount of energy. The momentum required is larger than 10 mini Brillouin zones (mBZs), where the exciton-phonon matrix element becomes negligibly small, as the overlap of the moir\'e exciton wavefunctions of the involved states is mostly localized in the first mBZ \cite{brem2020tunable}. In contrast,  optical phonons exhibit an energy of $\simeq 20-25$ meV that is closer to the energy condition required for the transition. The activation of this channel is explained in terms of the temperature-dependent broadening of the phonon-induced dephasing.  

 Given the energetic arguments presented above, the key quantities for the emergence of the bottleneck are the dispersion of the first excited and the ground state and the energy gap between them. Increasing the twist angle significantly modifies the dispersion of exciton bands towards a regular parabolic shape. At the same time the gap between the bands becomes smaller allowing for more scattering partners for both acoustic and optical phonon-driven scattering. A direct consequence is the decreasing importance of the relaxation  bottleneck at larger twist angles. The relaxations dynamics for 3$^\circ$ is shown and further discussed in the SI. 

The moir\'e exciton relaxation dynamics discussed so far in the momentum space has also a repercussion in the real space.  The ground state  X$_0$ is characterized by an s-type wave function, whereas the excited states X$_{1,2}$ are described by p-type wave functions \cite{brem2020tunable}. 
As depicted in \ref{fig:schematic}(b), the relaxation dynamics starts with free states, characterized by a spatially delocalized wavefunction. These relax to a localized equilibrium distribution in the ground state with a threefold s-type orbital that is maximally centered at each moir\'e potential minimum. In presence of a pronounced bottleneck, one has an equilibrium distribution with excitons occupying both the ground state and excited states. Translated in real space, this results  in a mixture of s-type and p-type orbitals around the moir\'e traps. As p-type states are characterized by a broader excitonic wavefunction than their s-type counterpart, one finds a larger excitonic wavefunction overlap between different spatial traps, resulting in an increase of the tunneling probability (more details can be found in the SI). As a result, the predicted relaxation bottleneck does not only lead to an enhanced lifetime, occupation, and emission intensity of excited states, but is also expected to give rise to a more efficient  propagation of moir\'e excitons. \\

\textbf{Conclusion}:
In this study, we have investigated the relaxation dynamics of interlayer excitons in a twisted MoSe$_2$-WSe$_2$ heterostructure exhibiting flat moir\'e bands. Based on a microscopic model including the twist-angle dependent moir\'e potential, we demonstrate the relaxation cascade of an initial hot distribution of interlayer excitons and identify a pronounced relaxation bottleneck at low temperatures and low twist angles. This drastically slows down the thermalization of excitons resulting in quasi-stationary exciton occupations far away from the Boltzmann distribution. A direct consequence is a higher occupation of excited exciton states explaining their larger emission compared to the ground state - in excellent qualitatively agreement with experimental  observations in  photoluminescence spectra of twisted TMD heterostructures.
Furthermore, we studied the temperature-dependent activation of the relaxation bottleneck, tracing back its microscopic origin to a combination of the energy separation and the flatness of the involved moir\'e exciton sub-bands. Overall, our study provides important microscopic insights into the exciton relaxation behaviour in presence of flat moir\'e bands in twisted van der Waals heterostructures.
\\

\textbf{Acknowledgements:} 
This project has received funding from Deutsche Forschungsgemeinschaft via CRC 1083 and the project 512604469. \\

\bibliography{references}

\end{document}


\title{Excitonic thermalization bottleneck in twisted TMDs heterostructures\\
\Large Supplementary Information}

\author{Giuseppe Meneghini}
\email{giuseppe.meneghini@physik.uni-marburg.de}
\affiliation{%
 Department of Physics, Philipps University of Marburg, 35037 Marburg, Germany}%

\author{Samuel Brem}

\affiliation{%
 Department of Physics, Philipps University of Marburg, 35037 Marburg, Germany}%

\author{Ermin Malic}
\affiliation{%
 Department of Physics, Philipps University of Marburg, 35037 Marburg, Germany}%

\maketitle
\thispagestyle{plain}

\section{Theoretical Approach}
\noindent
In this section we introduce a detailed derivation of the equations of motion shown in the main text. 

\subsection{Exciton bandstructure and dynamics}
\noindent
First, we introduce the change of basis to obtain the moiré Hamilton operator. The starting point is the Hamiltonian in second quantization describing a TMD bilayer system \cite{ovesen2019interlayer,brem2020tunable}, where we include the free exciton energy and the exciton-phonon interaction yielding
\begin{equation}\label{eq:ham_init}
H=\sum_{\mu {\bf Q}} \mathcal{E}^{\mu }_{{\bf Q}} X^{\mu\dagger}_{{\bf Q}} X^{\mu}_{{\bf Q}}+ \sum_{j{\bf Q}{\bf q} \mu\nu}{ \tilde{\Da}^{\mu\nu}_{j {\bf q} {\bf Q}} X^{\nu\dagger }_{{\bf Q + q}}X^{\mu}_{{\bf Q}} b_{j,{\bf q}} } + h.c. 
\end{equation}
Here, we use a valley-local approach including the exciton quantum number $\mu = (n^\mu, \zeta^\mu_e,\zeta^\mu_h,l^\mu_e,l^\mu_h)$, where $n$ describes the series of Rydberg-like states determining the relative electron-hole motion. Furthermore,  $\zeta_{e/h}$ and $l_{e/h} = 0,1$ correspond to the electron/hole valley and the layer index, respectively. We use also in the case of phonons a compound mode index $j = (\kappa_j,\zeta^{ph}_j,l^{ph}_j)$, where $\kappa$ is the phonon mode, $\zeta$ and $l$ phonon valley and layer respectively. Moreover, we have introduced  the exciton annihilation (creation) operators $X^{(\dagger)}$ and the exciton energy $\mathcal{E}^\mu_{\bf Q} = \hbar^2 {\bf Q}^2/(2 M_\mu) + E^g_\mu + E^b_\mu$ at the center-of-mass momentum ${\bf Q}$ with the mass $M^\mu =m^\mu_e+m^\mu_h$ ($m_{e/h}$ electron/hole mass). Here, $E^g_\mu$ corresponds to the energy gap between the valence and the conduction band and $E^b_\mu$ denotes the exciton binding energy, obtained from solving the Wannier equation.

For the exciton-phonon contribution, we have introduced  the exciton-phonon coupling element $\tilde{\Da}^{\nu\mu}_{j {\bf q} {\bf Q}}$ reading
\begin{align}
\begin{split}
\label{eq:exciton-phonon couplings}
\tilde{\Da}^{\mu\nu}_{j,{\bf q},{\bf Q}} &= D^{\zeta^e_\mu \zeta^e_\nu c}_{j,{\bf q}} \delta_{\zeta^h_\mu \zeta^h_\nu} \delta_{\zeta^e_\nu-\zeta^e_\mu, \zeta^{ph}_j} \delta_{l^e_\nu,l^{ph}_j} \delta_{l^e_\nu,l^e_\mu} \Fa^{\mu \nu}\left(  \frac{m^\nu_h}{M^\nu} \left[ {\bf q} + s_{\mu\nu}{\bf Q}  \right] \right)+\\[6pt]
    &- D^{\zeta^h_\mu \zeta^h_\nu v}_{j,{\bf q}} \delta_{\zeta^e_\mu \zeta^e_\nu} \delta_{\zeta^h_\nu-\zeta^h_\mu, \zeta^{ph}_j} \delta_{l^h_\nu,l^{ph}_j} \delta_{l^h_\nu,l^h_\mu} \Fa^{\mu \nu}\left(  - \frac{m^\nu_e}{M^\nu} \left[ {\bf q} + s_{\mu\nu}{\bf Q}  \right] \right).
\end{split}
\end{align}
Here, we use the subscript $ph$ to label phonon quantum numbers, and the terms $\delta_{\zeta^{e/h}_\mu \zeta^{e/h}_\nu}\delta_{\zeta^{h/e}_\nu-\zeta^{h/e}_\mu, \zeta^{ph}_j}$ fix the momentum conservation of the scattering process with the  phonon momentum $\tilde{\bf q} = \zeta^{ph}_j + {\bf q} $. In addition we have introduced the form factors $\Fa^{\mu \nu}\left( {\bf q} \right) = \sum_{\bf k}{ \psi^{\mu*}_{}({\bf k} )\psi^{\nu}_{}({\bf k +q} )}$ with the excitonic eigenfunction $ \psi^{\mu}({\bf k})$,  $s_{\mu\nu} = 1-M_\nu/M_\mu$, and $D^{\zeta^\lambda_m\zeta^\lambda_n\lambda}_{j,{\bf q}}$ as the electron/hole-phonon coupling element for TMD monolayers, taken from first-principle  calculations \cite{jin14}, yielding
\begin{align}
\begin{split}
    D^{\zeta^\lambda_m\zeta^\lambda_n\lambda}_{j,{\bf q}} &\approx \sqrt{\frac{\hbar}{2 \rho_{l^{ph}_j} A \Omega_{j{\bf q}} } } \Tilde{D} ^{\zeta^\lambda_m\zeta^\lambda_n\lambda}_{j,{\bf q}}\\[6pt]
    \text{with} \quad \Tilde{D} ^{\zeta^\lambda_m\zeta^\lambda_n\lambda}_{j,{\bf q}} &= 
    \begin{cases}
        \Tilde{D}^{\lambda}_\zeta {\bf q} \quad\text{if } \zeta^{\lambda}_m=\zeta^{\lambda}_n = \zeta \quad\text{and } \kappa_j =TA,LA\\[6pt]
        \Tilde{D}^{\lambda}_{\zeta^{\lambda}_m \zeta^{\lambda}_n} \quad\text{else }
    \end{cases}\\
    \text{and}\quad \Omega_{j{\bf q}} &= 
    \begin{cases}
        v_j {\bf q}  \quad\text{if } \kappa_j =TA,LA\\
        \Omega_{_j} \quad\text{else}.
    \end{cases}
\end{split}
\end{align}
Here, $\lambda = c,v$ corresponds to the electronic band index,  $TA,LA$ to the acoustic transversal and longitudinal phonon modes, and with the "else" we refer to the optical modes and intervalley contributions. Furthermore,  $A$ denotes the area of the system, $\rho_{l^{ph}_j}$ the surface mass density in the specific phonon layer $l^{ph}_j$, and $v_j$ the sound velocity in the TMD layer. 
For our study, we focus only on  low-energy excitations close to the ground state. Therefore, we can neglect the tunnelling-induced mixing of intra- and interlayer states, since the ground state in the investigated MoSe$_2$-WSe$_2$ heterostructure is a KK interlayer exciton and the electron/hole tunneling is negligible around the K valley \cite{hagel2021exciton}. In this context, we treat the effect of the twist angle as an effective potential, the so-called moiré potential reading \cite{brem2020tunable}
\begin{equation}
    V_M = \sum_{\bf Q \Tilde{g} g} \mathcal{M}_{\bf \Tilde{g} g} X^{\dagger}_{ \bf Q \Tilde{g}} X_{ \bf Q g}
\end{equation} 
where the exciton quantum number $\mu$ is fixed to be the exciton ground state and thus omitted in the following. Furthemore, ${\bf Q} \in 1^{st}$MBZ and ${\bf g} = s_1 {\bf G^M_1} + s_2 {\bf G^M_2}$ with ${\bf G^M_{1/2}}$ as the reciprocal moiré lattice vector and $s_{1/2}$ as integers (where we use MBZ to refer to the moiré Brillouin zone accounting for  the new periodicity arising from the moiré potential). The moiré matrix elements are defined as follows
\begin{align}
\begin{split}
    \mathcal{M}_{\bf \Tilde{g} g} &= \Theta \left( \delta_{s_1,\tilde{s_1}+(-1)^{l_e}} \delta_{s_2,\tilde{s_2}}  
    +\delta_{s_1,\tilde{s_1}}\delta_{s_2,\tilde{s_2}+(-1)^{l_e}}
    +\delta_{s_1,\tilde{s_1}+(-1)^{l_e}} \delta_{s_2,\tilde{s_2}+(-1)^{l_e}} \right) +\\
    &+ \Theta^* \left( \delta_{s_1,\tilde{s_1}-(-1)^{l_e}} \delta_{s_2,\tilde{s_2}}  
    +\delta_{s_1,\tilde{s_1}}\delta_{s_2,\tilde{s_2}-(-1)^{l_e}}
    +\delta_{s_1,\tilde{s_1}-(-1)^{l_e}} \delta_{s_2,\tilde{s_2}-(-1)^{l_e}} \right)
\end{split}
\end{align}
where $\Theta = v^c_{l_e} \Fa(\frac{m_h}{M} {\bf g_0}) - v^{v*}_{l_h} \Fa(\frac{m_e}{M} {\bf g_0})$ with $v^{c/v}_{l_{e/h}} = \gamma^{c/v}_1 + \gamma^{c/v}_2 e^{2\pi i /3}$ as the effective atomic potentials for the conduction and the valence bands in the neighbouring layer. The parameters $ \gamma^{c/v}_{1/2}$ are obtained from first-principle calculations, and ${\bf g_n} = \mathrm{C}^{n-1}_3({\bf G^1_1 - G^0_1})$, where $G^l_m$ refers to the $m =1,2$ reciprocal lattice vector for layer $l = 0,1$. 
By introducing the moiré potential and using the new periodicity of the system we can rewrite the free exciton Hamiltonian for the ground state in the presence of moiré potential in the following way \cite{brem2020tunable}
\begin{equation}
H_M = \sum_{{\bf Q g}} \mathcal{E}^{}_{{\bf Q g}} X^{\dagger}_{{\bf Q g}} X^{}_{{\bf Q g}}+ \sum_{\bf Q \Tilde{g} g} \mathcal{M}_{\bf \Tilde{g} g} X^{\dagger}_{ \bf Q \Tilde{g}} X_{ \bf Q g}
\end{equation}  
where we decompose the total center-of-mass momentum in terms of the new moiré reciprocal lattice vector  ${\bf \tilde{Q} = Q + g}$ with the new quantum number ${ \bf g}$. This Hamiltonian is diagonal for moiré excitons, i.e. $Y^{\eta (\dagger) }_{\bf Q} = \sum_{\bf g} \omega^{\eta (*)}_{\bf g}({\bf Q}) X^{(\dagger)}_{\bf Q g}$, when the momentum-mixing coefficients $\omega^{\eta (*)}_{\bf g}({\bf Q})$ fulfill the eigenvalue problem 
\begin{equation}
     \mathcal{E}^{}_{{\bf Q g}} \omega^{\eta}_{\bf g}({\bf Q}) + \sum_{\bf \tilde{g}} \mathcal{M}_{\bf \tilde{g} g} \omega^{\eta }_{\bf \tilde{g}}({\bf Q}) = E^\eta_{{\bf Q}} \omega^{\eta }_{\bf g}({\bf Q}).
\end{equation}
Using these states to perform a change of basis in the full Hamiltonian in Eq. (\ref{eq:ham_init}) leads us to the final Hamilton operator
\begin{equation}
    \label{eq:hamiltonian}
    \tilde{H} = \sum_{\eta}E^\eta_{ \bf Q} Y^{\eta \dagger}_{ \bf Q} Y^\eta_{ \bf Q} + \sum_{\substack{\eta \xi j \\{\bf Q Q' g} }} \Tilde{\mathcal{D}}^{\eta \xi j}_{{\bf Q Q' g }} Y^{\xi \dagger}_{{\bf Q'}} Y^\eta _{{\bf Q}} b^j_{{\bf Q'-Q+  g}} + h.c.
\end{equation}
where the moiré exciton-phonon coupling elements are defined as follows
\begin{equation}
    \tilde{\mathcal{D}}^{\eta \xi j}_{{\bf Q Q' g }} =  \sum_{\bf g' \tilde{g}} \tilde{\mathcal{D}}^{}_{j {\bf Q'-Q+g, Q }}  \omega^{\eta *}_{\bf \tilde{g}}({\bf Q})\omega^{\xi }_{\bf g'}({\bf Q'}) \delta_{\bf g, g'-\tilde{g}}.
\end{equation}
These are expressed in terms of exciton-phonon coupling elements $\tilde{\mathcal{D}}^{}_{j {\bf Q'-Q+g, Q }}$ defined in Eq. (\ref{eq:exciton-phonon couplings}). 

With the full Hamiltonian, we solve the Heisenberg equation of motion for moiré excitons occupation $N^\eta_{\bf Q} = \left< Y^{\eta \dagger} _{{\bf Q}} Y^\eta _{{\bf Q}} \right>$, truncating the Martin-Schwinger hierarchy using a second-order Born-Markov approximation, obtaining \cite{brem2018exciton, selig2018dark}
\begin{equation}
    \Dot{N^\eta_{\bf Q}} = \sum_{\xi \bf Q'} \left( \mathcal{W}^{ \xi\eta }_{ \bf Q' Q} N^\xi_{\bf Q'} - \mathcal{W}^{ \eta \xi}_{ \bf Q Q'} N^\eta_{\bf Q} \right).
\end{equation}
The phonon-mediated scattering tensor, including emission and absorption processes ($\pm$), reads
\begin{equation}
    \label{eq:scattering_rate}
     \mathcal{W}^{ \eta \xi}_{ \bf Q Q'} = \sum_{j \pm {\bf g}}  \left| \Tilde{\mathcal{D}}^{\eta \xi j}_{{\bf Q' Q g}}\right|^2 \left( \frac{1}{2}\pm \frac{1}{2} + n^B_{j {\bf Q'-Q+g}}  \right) \delta\left( \Delta^{\eta \xi \pm}_{\bf Q Q' g} \right)
\end{equation}  
with the energy conservation $\Delta^{\eta \xi \pm}_{\bf Q Q' g} = E^\xi_{ \bf Q'} - E^\eta_{ \bf Q} \pm \Omega_{j \bf Q' -Q +g}$, where $ n^B_{j {\bf Q'-Q+g}}$ is the Bose-Einstein distribution for phonons with the mode $j$, the momentum ${\bf Q'-Q+g}$, and the energy $\Omega_{j \bf Q' -Q +g}$.
Using the scattering tensor we can define the density of in-scattering states for the specific moiré exciton state $\eta$, yielding
\begin{equation}
    \mathrm{DO\Gamma}^\eta_{in} (E)= \sum_{\xi {\bf Q'}} \mathcal{W}^{\xi\eta }_{ \bf  Q', Q = 0} \delta\left( E - E^\xi_{\bf Q'}\right).
\end{equation}
This quantifies the density of states contributing the most to the influx of excitons to a specific state $\eta$.
Additionally, an important quantity to monitor how the optical selection rules of interlayer excitons are influenced by the moiré potential is the absorption reading \cite{brem2020tunable}
\begin{align}
\begin{split}
    \alpha(E) &= \sum_{\eta \sigma} \left| \gamma^\eta_\sigma \right|^2 \mathcal{L}_{\Gamma^\eta}(E^\eta_{\bf Q = 0} - E)  \\
    \text{with}\quad \gamma^\eta_\sigma &= \sum_{\bf g} \omega^\eta_{\bf g}({\bf Q = 0}) \tilde{\gamma}_\sigma({\bf g})\\
    \text{with}\quad \tilde{\gamma}_\sigma({\bf q}) &= \Omega_L \sum^2_{n=0} e^{- i 2 n \pi/3 } \delta_{\bf q,g_n} \frac{\mathcal{C}^{n-1}_3({\bf K}_l)}{| {\bf K}_l|}\cdot{\bf e_\sigma}
\end{split}
\end{align}
with $\Omega_L \propto \sum_{\bf k} \psi_L({\bf k})$ and $L = (l_e,l_h)$.  All equations  are evaluated at ${\bf Q = 0}$ corresponding to  the minimum of the moirè exciton dispersion. This means that in the case of intralayer excitons the zero would be exactly at ${\bf Q = 0}$, since the relative momentum displacement of electrons and holes is zero. In contrast,  for interlayer excitons the equations are evaluated at $Q = \kappa$ (K point of the MBZ), reflecting the mismatch of the Brilloin zones of the two layers.

\subsection{Self-consistent dephasing rate, photoluminescence intensity, and relaxation time}
\begin{wrapfigure}[19]{r}{0.4\textwidth}
\vspace{-38pt}
  \begin{center}
    \includegraphics[width=0.38\textwidth]{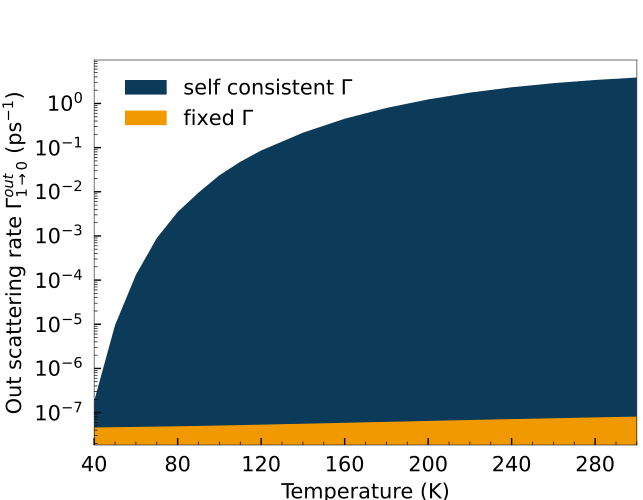}
  \end{center}
  \vspace{-4pt}
  \captionsetup{margin=0.1cm,justification=justified}
  \caption{Comparison of the temperature dependent out-scattering rate from the first excited states to the ground state using the self-consistent approach and a fixed width of the Cauchy/Lorentz distribution in Eq. (\ref{eq:selfGamma}).}
    \label{fig:compar}
\end{wrapfigure}
\noindent
Equation (\ref{eq:scattering_rate}) is derived by using a second-order Born-Markov approximation \cite{haug2009quantum, thranhardt2000quantum,selig2018dark,brem2018exciton}, thus obtaining fixed resonances from the energy conservation described by the delta function. A more general approach can be used to include a self-consistent temperature-dependent broadening. The latter can be obtained by continuing the correlation expansion taking into account two-particle correlations and considering only the imaginary part of the self-energy (neglecting polaron renormalization terms) \cite{rossi2002theory}. 
Considering only the phonon contribution, we can define the dephasing rate as follows  
\begin{equation}
\label{eq:selfGamma}
\footnotesize
    \Gamma^{\eta \xi}_{\bf Q} = \frac{\hbar}{2} \sum_{j \pm {\bf Q' g} }  \left| \Tilde{\mathcal{D}}^{\eta \xi j}_{{\bf Q' Q g}}\right|^2 \left( \frac{1}{2}\pm \frac{1}{2} + n^B_{j {\bf Q'-Q+g}}  \right) \mathcal{L}_{\Gamma^\eta_{\bf Q} + \Gamma^\xi_{\bf Q'}}\left( \Delta^{\eta \xi \pm}_{\bf Q Q' g} \right)
\end{equation}
resulting in a system of coupled equations that can be solved self-consistently. Here, $\mathcal{L}_\Gamma$ corresponds to  the Cauchy/Lorentz distribution with the width $\Gamma$.
We directly compare a self-consistent calculation  of the out-scattering rate from the first excited states to the ground state  including the temperature-dependent broadening with a calculation based on a fixed width $\Gamma=1$ meV of the Cauchy/Lorentz distribution (chosen to be close to the value of the lowest temperature of the self-consistent treatment), cf. Fig. \ref{fig:compar}. We find an out-scattering that is by orders of magnitude more pronounced at enhanced temperatures for the self-consistent treatment reflecting the softening of the strict energy conservation. 
Finally, we use Eq. (\ref{eq:selfGamma}) to determine  the time- and energy-dependent photoluminescence intensity $I_{PL}(E,t)$ for moiré excitons \cite{koch2006semiconductor,brem2020tunable} reading
\begin{equation}
    I_{PL}(E,t) \propto \sum_{\eta,\sigma} \left|\gamma^\eta_{\sigma}\right|^2 N^\eta_{\bf Q = 0} \mathcal{L}_{\Gamma^\eta}(E^\eta_{\bf Q = 0} - E) . 
\end{equation}
In addition, we define the relaxation time $\tau_{\eta \rightarrow \xi}$ quantifying the scattering from the moiré exciton state $\eta$ to the state $\xi$ yielding 
\begin{equation}
    \tau_{\eta \rightarrow \xi} = \frac{1}{\Gamma^{\eta \xi}_{\bf Q = 0}}.
\end{equation}
We use this equation in the main text to characterize the relaxation bottleneck effect and study its temperature dependence.

\section*{Dependence on the initial excitation conditions}
\begin{figure}
  \centering
  \includegraphics[width=\columnwidth]{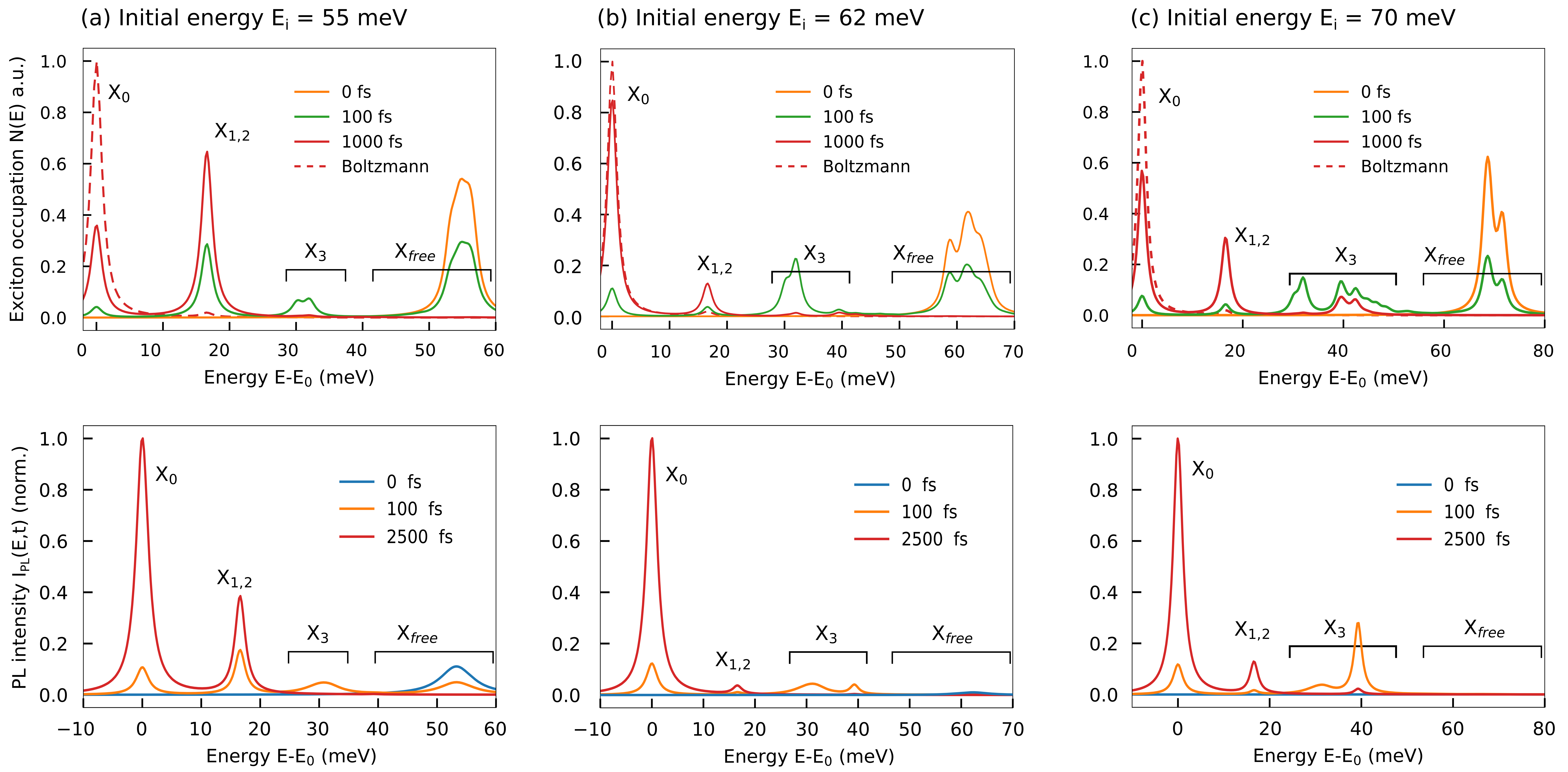}
  \caption{Study of the moiré exciton dynamics at 40 K for different initial conditions. We initialize the system with a uniform energy distribution of excitons centered at (a) $E_i = $55 meV, (b) $E_i = $62 meV and (c) $E_i = $70 meV. In the top row we show energy-resolved and momentum-integrated exciton occupation at different time cuts with the red dashed line corresponding to the Boltzmann distribution. In the bottom row, we show photoluminescence spectra as a function of energy  at different fixed time cuts.}
  \label{fig:initial_condition}
\end{figure}

\noindent
In the main text we have briefly discussed that the specific ratio of exciton occupation distributed between the ground state and the first excited states is strongly influenced by the initial excitation condition, more specifically on the excitation energy window. We show in this section that, although the specific output of the dynamics can be dependent on the initial condition, the general physical result is still valid, i.e. the final distribution of excitons at low temperatures and low twist angles deviates from a Boltzmann distribution due to the emergence of a relaxation bottleneck effect. 
In particular, we vary the initial exciton energy distribution $E_i = 55,62,70$ meV, and calculate the final  steady-state exciton distribution, cf. Fig.  \ref{fig:initial_condition}.
 The top row in Fig.  \ref{fig:initial_condition} shows different time cuts of the momentum-integrated energy-resolved moiré exciton occupation. A common aspect of the exciton dynamics  is the very fast initial dissipation of energy that brings the initial distribution of excitons to intermediate states. Focusing on the 1000 fs time cut, we see that the dynamics has already reached a stationary solution, and we find a clear deviation states from the thermal Boltzmann distribution at 40 K (dashed line). The quantitative percentage of deviation depends on the initial condition: if excitons can emit optical phonons during the dissipation process to reach efficiently intermediate states (X$_3$ states), the occupation of the ground state becomes greater than the first excited states. This is the case for initial exciton occupations at 62 and 70 meV and is explained in terms of the X$_3$ states scattering efficiently to the ground state, cf. Fig. \ref{fig:initial_condition} (b,c). In contrast,  states at energies around 55 meV scatter more efficiently to the first excited states, cf. Fig. \ref{fig:initial_condition} (a). The larger occupation of the ground state is accompanied by a brighter response of X$_0$ in PL spectra, cf. the lower panels in Fig. \ref{fig:initial_condition}. 

\begin{figure}[b!]
  \centering
  \includegraphics[width=.6\columnwidth]{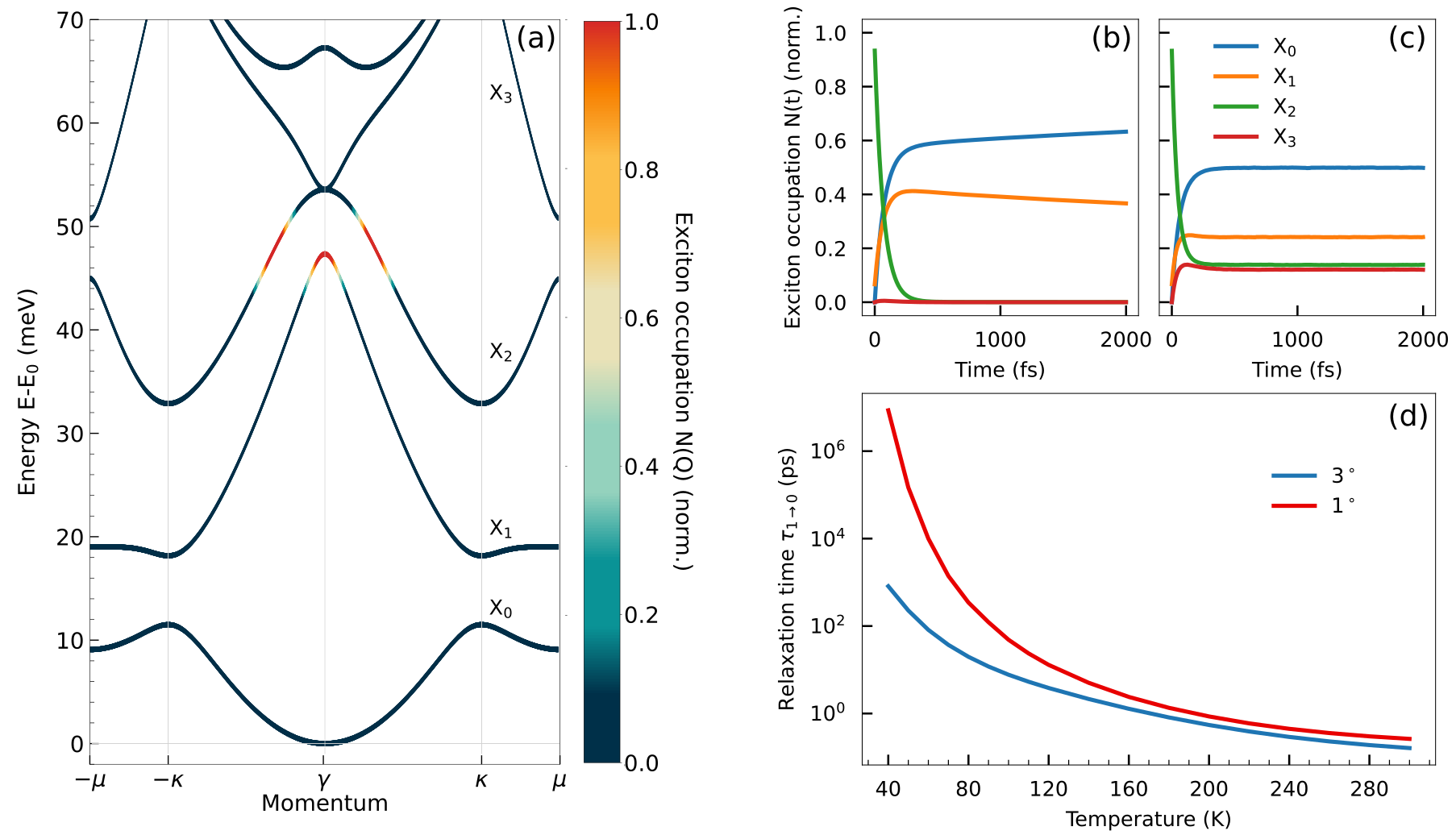}
  \caption{Relaxation dynamics at the larger twist angle of $3^\circ$.  (a) Moiré exciton band structure, where the initial momentum-dependent occupation is highlighted with a color scheme.  We plot the momentum-integrated and time-dependent exciton occupation for each band relevant for the relaxation process at (b) 40 K and (c) 300 K. (d) Direct comparison of the temperature-dependent relaxation time at 1$^\circ$ and 3$^\circ$  showing that the X$_1 \rightarrow$ X$_0$ transition is slow even for 3$^\circ$, but still several orders of magnitude faster than at 1$^\circ$, where the flat bands strongly restrict the number of possible scattering states. }
  \label{fig:3deg}
\end{figure}

\section*{Twist angle dependence of the relaxation bottleneck}
\noindent
In the main text we focus on analyzing the small twist angle regime, pointing out that the appearance of flat bands has a key role for the emergence of the relaxation bottleneck during the exciton thermalization process. Here, we show a large twist angle case ($\simeq 3^\circ$), where the exciton bands are not flat any longer, cf. Fig. \ref{fig:3deg}(a). To be consistent with the study in the main text, we choose an initial condition for the excitation in the same energy window ($\simeq$ 48 meV from the ground state minimum). Following the time evolution and conducting the same temperature-dependent study, we determine the relaxation time $\tau_{1\rightarrow0}$ that is now strongly momentum- dependent. We calculate  $\tau_{1\rightarrow0}$ at the corners of the mini Brillouin zone, where the first excited state exhibits a minimum. We show the exciton dynamics at 40 K and 300 K in  Figs. \ref{fig:3deg}(b)-(c). We directly compare $\tau_{1\rightarrow0}$ in the small (red line) and the large (blue line) twist angle regime, cf.  Fig.\ref{fig:3deg}(d). The largest difference in the relaxation time is found at low temperatures. The slowed-down relaxation process at 1$^\circ$  can be traced back to  flat exciton bands and the restricted scattering efficiency due to the energy conservation. This is pronounced, in particular, at low temperatures, where the broadening of states is small and thus a strict energy conservation needs to be fulfilled. The effect is much less pronounced at 3$^\circ$ exhibiting parabolic bands, where the number of possible scattering partners is much higher than in the case of flat bands at 1$^\circ$. As the temperature increases the relaxation time at both twist angles starts to merge leading to a comparable relaxation time at room temperature. Overall, we can conclude that the key ingredient for the emergence of the relaxation bottleneck is the peculiar flat  bandstructure of moiré excitons.

\section*{Moiré exciton distribution in real space}

\begin{figure}[t!]
  \centering
  \includegraphics[width=.7\columnwidth]{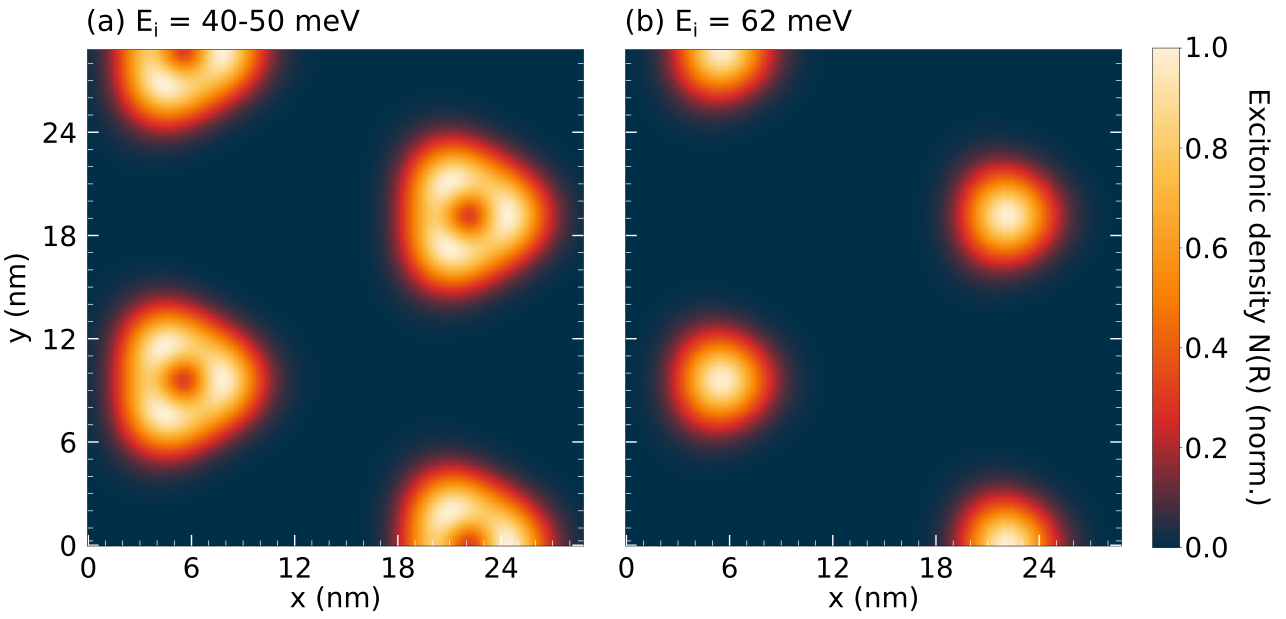}
  \caption{Equilibrium moiré exciton distribution in real space, highlighting the different situation in the case of (a) highly occupied excited states with a p-type orbital character (obtained for initial exciton occupations in the energy window of 40-50 meV) and (b) highly occupied ground state with an s-type orbital character (obtained
  for an initial exciton occupation around 62 meV).}
  \label{fig:Real_space_comp}
\end{figure}
\noindent
In the main text we briefly state that the relaxation bottleneck also influences the real space distribution of excitons - depending on which state is mostly occupied at equilibrium. This is in particular valid for  small twist angles, where the ground state and the first excited states exhibit very different excitonic wavefunctions. While the ground state wavefunction is of an s-type character, the first excited states are rather p-type-like. We investigate the change in the exciton distribution in the real space for a twist angle of 1$^\circ$ (i) for an initial exciton occupation in the energy window of 40-50 meV resulting in highly occupied excited states (Fig. \ref{fig:Real_space_comp}(a)) and (ii) for an initial exciton occupation around 62 meV resulting in a highly occupied ground state (Fig. \ref{fig:Real_space_comp}(b)). 
In the first case, we find that excitons have a p-type shape around each moiré trap in real space, while in the second case the exciton distribution  has the characteristic s-type shape centered at each moiré trap. As p-type orbitals have a broader profile in real space, this results in a larger overlap of the excitonic wavefunctions of neighbouring traps, thus affecting the tunneling rate between them.

\bibliography{references}